%% file: paper.tex
\documentclass[copyright,creativecommons]{eptcs}
\providecommand{\event}{ACL2 2013} 
\usepackage{breakurl}             

\usepackage{graphicx}
\usepackage{amsmath}

{\bfseries}{}
\newtheorem{example}{Example}{\bfseries}{}
{\bfseries}{}

\input{common.tex}

\dutch

\title{A formalisation of \xmas}
\author{Bernard van Gastel \qquad Julien Schmaltz
\institute{Open University of the Netherlands}
\email{\{Bernard.vanGastel, Julien.Schmaltz\}@ou.nl}
}
\def\titlerunning{A formalisation of \xmas}
\def\authorrunning{B. van Gastel and J. Schmaltz}

\begin{document}
\maketitle

\begin{abstract}
Communication fabrics play a key role in the correctness and performance of modern multi-core processors and systems-on-chip. To enable formal verification, a recent trend is to use high-level micro-architectural models to capture designers' intent about the communication and processing of messages. Intel proposed the \xmas language to support the formal definition of executable specifications of micro-architectures. We formalise the semantics of \xmas in \acl. Our formalisation represents the computation of the values of all wires of a design. Our main function computes a set of possible routing targets for each message and whether a message can make progress according to the current network state. We prove several properties on the semantics, including termination, non-emptiness of routing, and correctness of progress conditions. Our current effort focuses on a basic subset of the entire \xmas language, which includes queues, functions, and switches.
\end{abstract}

\section{Introduction}

Today's computing devices -- e.g. laptops, servers, embedded systems -- integrate a large number of processing and memory elements. 
In this realm of multi-core processors and multi-processors Systems-on-Chip, the design of the communication architecture 
plays a key role in the overall correctness and performance of a system. When the number of integrated components grows, 
complex Networks-on-Chip -- also called communication fabrics -- replace standard bus architectures~\cite{dally01,benini02}. As for all elements of the design, 
these complex integrated networks need to be verified. This is a real challenge. The concurrent nature of message transfers makes the application 
of simulation techniques difficult. Communication fabrics are large structures characterised by a large number of queues and distributed control. 
These network architectures are also defined by a large number of parameters, e.g. the size of messages or queues, the type of flow control or routing 
algorithm. All these aspects constitute challenges for the application of formal verification techniques.

A recent trend has been to propose the use of high-level micro-architectural models to ease the application of formal methods to on-chip 
interconnects. Intel recently proposed \xmas -- eXecutable Micro-Architecture Specifications -- to capture the intent of architects~\cite{chatterjee10}.
High-level \xmas models can be used to extract invariants, which are then given to a hardware model checker to perform verification 
of hardware designs~\cite{gotmanov11,chatterjee12}. An advantage of these techniques is to apply to the hardware description. 
An issue is that an \xmas model is required. Also,
there should be some correspondence between the \xmas model and its hardware implementation.  Another direction is to perform 
validation at the level of the \xmas model. Recent studies have described the verification of deadlock freedom of large micro-architectural 
descriptions~\cite{verbeekschmaltz:fmcad11} and initial efforts about proving end-to-end latency bounds~\cite{holcomb2012}. 
If scalability seems to be a clear advantage of this abstract layer, 
an important issue is the transfer of properties proven at the \xmas level to the hardware implementation. 
This issue still constitutes an open question. 
The use of \xmas models to improve hardware verification has only be applied to parts of communication fabrics. Network-wide verification is also an open question. 

Leaving the open question of relating \xmas models to hardware descriptions aside, our long-term objective is 
to support the efficient verification of detailed micro-architectural models of communication fabrics. 
Our approach is to define a set of proof obligations such that any \xmas network satisfying these proof 
obligations is correct. 
As correctness criterion, we consider the notion of correctness defined in the \genoc environment. 
 \genoc~\cite{Schmaltz:2008fp,borrione09,Verbeek:2012cv} is a specification and validation environment for 
communication network architectures. It has been entirely formalised in \acl.
The \genoc theory defines a general notion of correctness for communication network
architectures. This notion -- coined \emph{productivity}~\cite{freekthesis} -- states that all messages eventually gain 
access to the network and eventually reach their 
expected destination. The \genoc theory also identifies essential properties of each constituent of a network. 
These properties -- called proof obligations -- are sufficient to ensure productivity. 
Our approach is to instantiate these proof obligations for an arbitrary \xmas network. 
To achieve this objective, we need to encode the \xmas semantics as an instance of the \genoc 
theory. This encoding should reveal the proof obligations that are sufficient to ensure productivity of an arbitrary \xmas network. 
A major challenge is that the latest version of the theory is not rich enough to express the
xMAS semantics. Of interest to this paper is the issue that, currently, routing in \genoc only depends on the destination and the current
position of a message. In xMAS, as it will be explained later in the paper, messages are routed according to their content.
For instance, routing decisions depend on the type of messages, e.g., request or responses. 
The contribution of this paper is an important step towards the extension of \genoc and its instantiation for \xmas.
The important step presented in this paper consists in the formalisation of a core subset of the \xmas language in the 
context of the \genoc formalism. 

The \xmas language consists of eight primitives with well-defined semantics (see next section for more details). 
Our formalisation considers 
 five of them. These primitives are primarily concerned with routing messages and do not consider arbitration and synchronisation 
 between messages. 
 We describe our formal representation of an \xmas network in Section~\ref{sec:xmasrep}. 
 In Section~\ref{sec:semantics}, we define a function representing the semantics of the \xmas primitives. 
 This function mutually recursively computes the values of all (interdependent) signals, in such a way that the results 
 can be integrated in the \genoc environment.  
 We prove several properties on this function, including non-emptiness of routing and correctness of flow-control 
 decisions. The conclusion (Section~\ref{sec:conclusion}) relates these properties to the axioms of \genoc and sketches further
 research directions. 

\section{Background: \xmas and \genoc}

\subsection{The \xmas language}
\label{sec:xmas}

\begin{figure}[t]
  \centering
  \includegraphics[width=\textwidth]{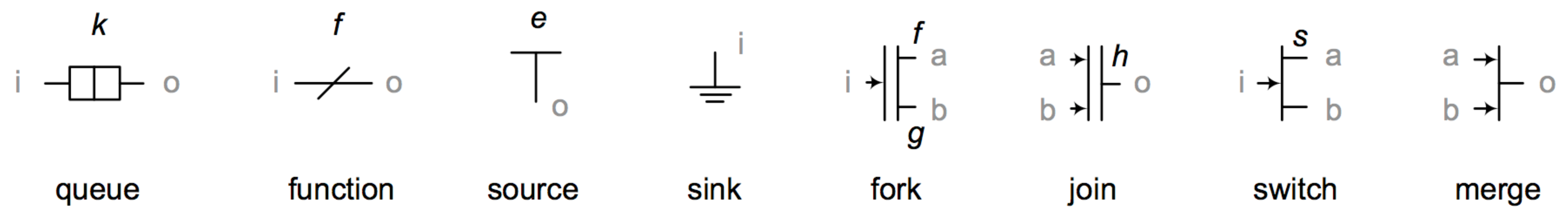}
  \caption{Eight primitives of the \xmas language. Italicised letters indicate parameters. Grey letters indicate ports.}
  \label{fig:xmas-prim}
\end{figure}

An \xmas model is a network of primitives connected via typed \emph{channels}. 
A channel is connected to an \emph{initiator} and a \emph{target} primitive.
A channel is composed of three signals.
Channel signal $\mathit{c.irdy}$ indicates whether the initiator is ready to write to channel $c$.
Channel signal $\mathit{c.trdy}$ indicates whether the target is ready to read from channel $c$.
Channel signal $\mathit{c.data}$ contains data that are transferred from the initiator output to the 
target input if and only if both signals $\mathit{c.irdy}$ and $\mathit{c.trdy}$ are set to true. 
Figure~\ref{fig:xmas-prim} shows the eight primitives of the \xmas language. 
A \emph{function} primitive manipulates data. Its parameter is a function that produces an outgoing packet 
from an incoming packet. Typically, functions are used to convert packet types and represent message dependencies
inside the fabric or in the model of the environment. 
A \emph{fork} duplicates an incoming packet to its two outputs. Such a transfer takes place if and only if the input is ready to send 
and the two outputs are both ready to read. 
A \emph{join} is the dual of a fork. The function parameter determines how the two incoming packets are merged. 
A transfer takes place if and only if the two inputs are ready to send and the output is ready to read. 
A \emph{switch} uses its function parameter to determine to which output an incoming packet must be routed. 
A \emph{merge} is an arbiter. It grants its output to one of its inputs. 
The arbitration policy is a parameter of the merge. For our purpose, 
only fair arbitration policies are supported. We abstract away from the details of the policy.
A \emph{queue} stores data. 
Messages are non-deterministically produced and consumed at \emph{sources} and \emph{sinks}. 
Sources and sinks are fair, i.e., packets are eventually created or consumed. A source 
or sink may process multiple packet types. 

\begin{figure}[h]
\centering
\includegraphics[scale=0.2]{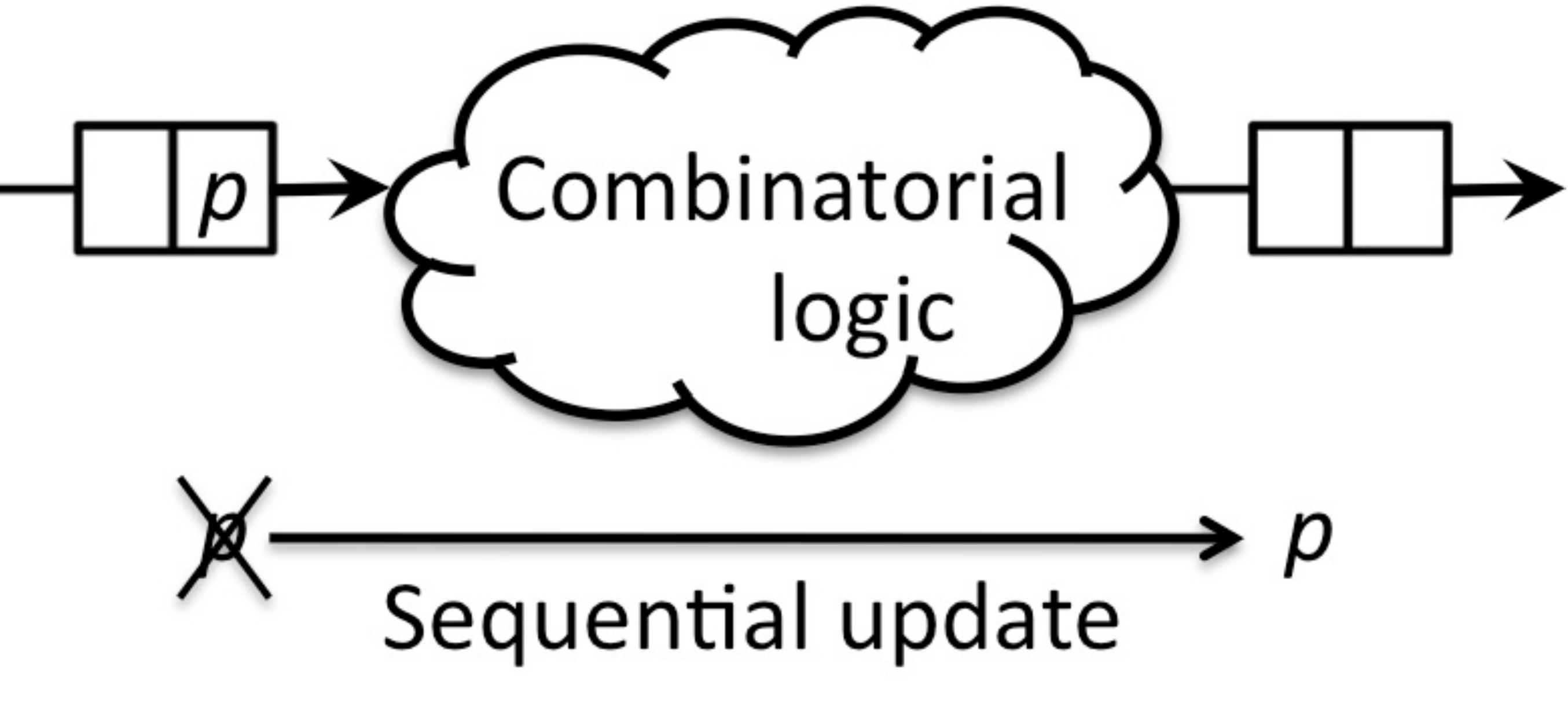}
\caption{Sequential updates regulated by combinatorial primitives}
\label{fig:execution}
\end{figure}

The execution semantics of an \xmas network consists in a combinatorial and a sequential part (Fig.~\ref{fig:execution}).
The combinatorial part updates the values of channel signals. The sequential part is the synchronous update of all 
queues according to the values of the channel signals. A simulation cycle consists of a combinatorial and a 
sequential update. A sequential update only concerns queues, sinks,
and sources. We denote these primitives as \emph{sequential primitives}. Other primitives are denoted as
\emph{combinatorial}. The semantics are only well defined if there are no combinatorial cycles.

For each output channel $o$, signal $\irdy$ is set to true if the
primitive can transmit a packet towards channel $o$, i.e., channel $o$
is ready to transmit to its target. 
For each input channel $i$, signal $\trdy$ is set to true if the
primitive can accept a packet from input channel $i$, i.e., the target
of channel $i$ is ready to receive.
In a sequential primitive, the values of output signals depend on
the values of the input signals and an internal state. Queues accept
packets only when they are not full. A source or a sink produces or
consumes a packet according to an internal oracle modelling
non-determinism.

\begin{figure}[h]
\centering
\includegraphics[scale=0.2]{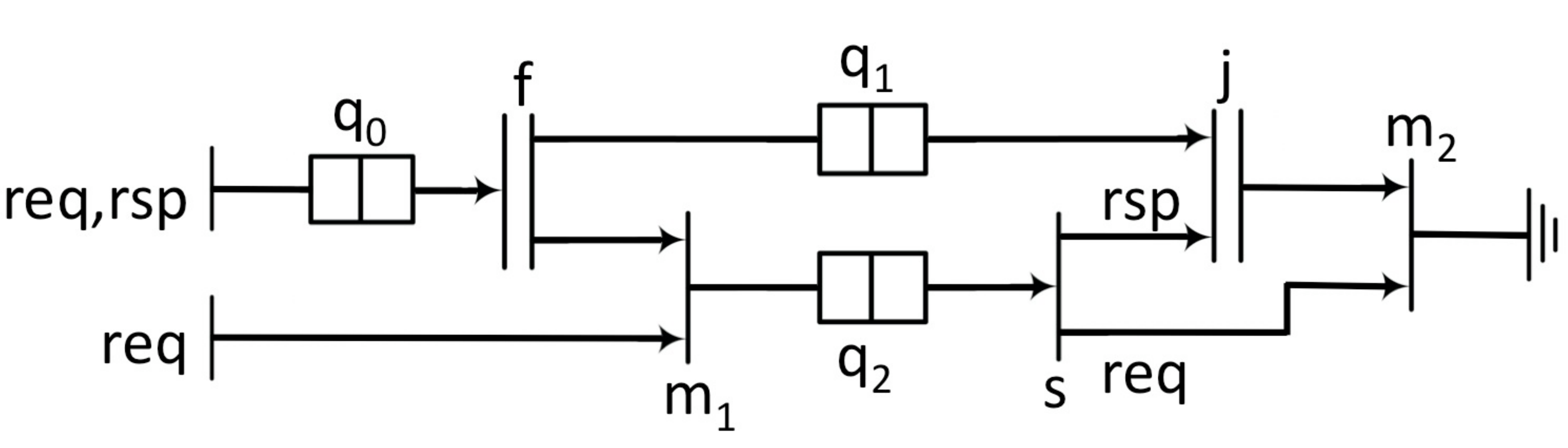}
\caption{Microarchitectural model}
\label{fig:xmas_ex}
\end{figure}

\begin{example}\label{ex:xmas}
  Consider the \xmas model in Fig.~\ref{fig:xmas_ex}. 
  Assume two request packets are injected in queue $q_0$ and the
  other source remains silent. At the fork, the combinatorial
  update propagates to queues $q_1$ and $q_2$ a positive
  $\mathit{irdy}$ signal. As queues $q_1$ and $q_2$ are ready to
  receive, a positive $\mathit{trdy}$ is propagated back to $q_0$.
  In the next sequential update, one request packet is moved to
  queues $q_1$ and $q_2$. The remaining packet will be moved in the
  next simulation cycle. Requests in $q_2$ are routed to the sink and
  are eventually consumed. The requests in $q_1$ are blocked as no
  response can arrive at the join. Indeed, responses injected in $q_0$
  are blocked by the requests in $q_1$ if $q_1$ is full. 
\end{example}

\subsection{The \genoc environment}
\label{sec:genoc}

\begin{figure}[htbp]
  \centering
    \includegraphics{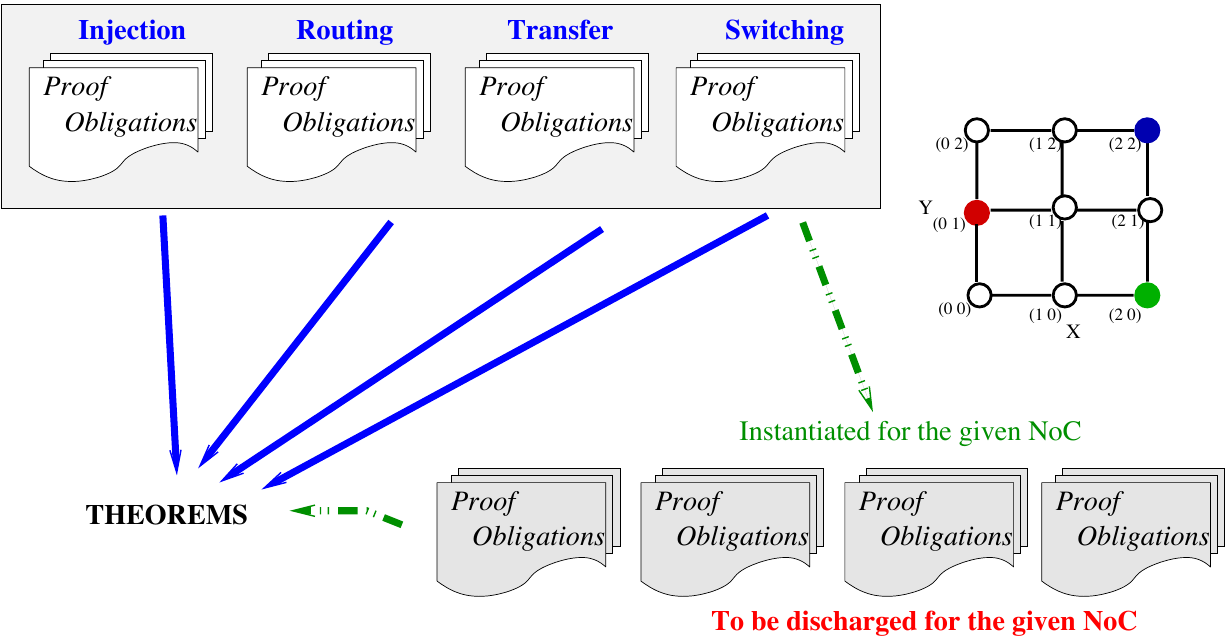}
  \caption{The \genoc methodology}
  \label{fig:genoc-overview}
  \todo{old figure, update}
\end{figure}

The \genoc approach provides an efficient specification and validation environment for 
abstract and parametric descriptions of NoCs communication infrastructures (see Figure~\ref{fig:genoc-overview}).
The \genoc model is a function representing 
the interaction between essential constituents -- e.g., routing function, switching policy -- common to a large class of network architectures. 
These constituents are not given a concrete definition but only characterised by a set of properties, called \emph{proof obligations}. 
The particular definitions of these constituents are parameters of the verification environment. The \genoc model is a meta-model of all 
particular architectures satisfying these proof obligations. The  correctness of the \genoc model is expressed as a theorem,
the proof of which directly follows from the proof obligations. Hence, once the proof obligations 
have been discharged for a particular architecture, it automatically follows that this architecture satisfies the global theorem. 
This generic aspect of \genoc is key. It reduces the verification to discharging proof obligations local to each constituent. 
This generic aspect also provides a compositional approach. Verified instances of the constituents can easily be re-used. 

  Function \genoc is the composition of the following five functions, representing 
  essential constituents of any network architecture:
  \begin{itemize}
  \item the \emph{injection method} ($I$) determines which pending messages can access the network;
  \item the \emph{routing function} ($R$) determines the next hop(s) from the current position and the destination of messages;
  \item the \emph{ordering function} ($O$) resolves arbitration conflicts;
  \item the \emph{transfer decision} ($T$) decides if a message can advance towards a next hop.
  \item the \emph{switching policy} ($S$) uses the routing function and the transfer decision to advance messages.
  \end{itemize}
  Function GeNoC combines these functions to form a network simulator. 
  It basically corresponds to the following function composition:
  
  \[ \genoc = S \circ T \circ O \circ R \circ I \]
  
  Function GeNoC takes a network state as input and simulates the NoC architecture until either 
  all communications have been processed or a deadlock has been reached. It 
  returns the final value of the network state. 
  Our objective is to encode the \xmas semantics in \genoc. The switching function directly 
  corresponds to the synchronous update of all queues. The ordering function relates 
  to the merge primitive. The injection function relates to sources. Functions transfer and routing are 
  the  parts of \genoc relevant to this paper. The other functions will not be discussed any further. 
  
  The most important proof obligations defined on functions transfer and routing are the following:
  \begin{itemize}
  \item {\bf There is always one routing target. }This means that routing must always propose at least one 
  possible next hop for a message. It is up to the transfer function to decide if a message can move to this destination.
  \item {\bf All routing targets are valid resources.} Resources correspond to elements which can inject, evacuate, or store messages. 
  This proof obligation ensure that routing only occurs from resources to resources. 
  \item {\bf Transfer is a subset of routing. }The purpose of the transfer function is to decide which messages 
  can actually move to their next hop. This proof obligation ensures consistency between the routing and the transfer phase.
  \item {\bf All possible transfers are directed to available resources.} This proof obligation ensures the correctness of the transfer phase. 
  Typically, a resource is available if it has space to store a message, e.g. a buffer or a queue. Correctness here means that the 
  transfer phase allows a transfer to a resource only if this resource has available space to store a message. 
  \end{itemize}
 
 In the remainder of this paper, we show the formalisation of functions routing and transfer for the \xmas language.

\section{Formal representation of an \xmas network}
\label{sec:xmasrep}

An \xmas network is formally represented as a list of channels and a list of components. A component is a structure
made of a type, inputs, outputs, and a field.  The constant {\tt componenttypes} currently includes 
the types queue, switch, source, sink, and function. The constant {\tt resourcetypes} is a subset of {\tt componenttypes} and include the three component types that can contain messages: queue, source and sink.

\begin{lstlisting}[numbers=none]
(defstructure component type ins outs field
  (:options (:representation :list)
  (:assert (and (member-equal type (componenttypes))))))
\end{lstlisting}

The {\tt field} part of the structure is used to store the function parameter of a component. Such a function 
is stored as an alist. Each key of the alist corresponds to a possible input argument of the function and 
stores the corresponding result. The following {\tt defun} event defines function application. 

\begin{lstlisting}[numbers=none]
(defun apply-field (n component parameter)
  (let ((func (nth n (component-field component))))
        (cdr (assoc parameter func))))
\end{lstlisting}

For instance, a component of type switch has only one function parameter. The following function 
computes the result of this parameter, but if the input is the special symbol \func{'error}, it preserves this symbol:

\begin{lstlisting}[numbers=none]
(defun xmas-switch-function (component data)
  (if (equal data 'error)
    'error
    (apply-field 0 component data)))
\end{lstlisting}

A channel is simply a structure composed of an initiator component and a target component. 

\begin{lstlisting}[numbers=none]
(defstructure channel init target
  (:options (:representation :list)))
\end{lstlisting}

There are a number of useful accessor functions for components, as both the {\tt ins} and {\tt outs} in the components structures are channel references. 
The representation is an integer serving as an index in a component structure. Function \func{get-in-channel} 
accesses input channels of a component. For instance, the call \func{(get-in-channel component i ntk)}
produces the i'th input channel of component \func{component} of an xMAS network \func{ntk}.
Likewise function \func{get-out-channel} accesses output channels of a component.

Similar to accessor functions of components, the following accessor functions
are defined for channels: \func{get-init-component} and \func{get-target-component}. 
The former returns the initiator component of a channel. The later returns the target component 
of a channel. 
Both take two arguments: a channel and an xMAS network.

There are several properties needed to ensure that an \xmas network is well-formed. 
A first property ensures uniqueness of the target component of a channel.
This is expressed in \acl using the following {\tt defun-sk}, which states 
that for all input ports of a component, all input channels have only that precise 
component as target component:

\begin{lstlisting}[numbers=none]
(defun-sk component-to-channel-invertible-ins (ntk)
  (forall (component i)
          (implies (and (componentp component 
                                    (len (xmasnetwork-channels ntk)))
                        (< i (len (component-ins component))))
                   (equal (get-target-component (get-in-channel component i ntk) ntk)
                          component))))                          
\end{lstlisting}

Similarly, we need to express uniqueness of the initiator component of each 
channel. The following {\tt defun-sk} states that for all output ports of a component, 
all output channels have only that component as initiator. 

\begin{lstlisting}[numbers=none]
(defun-sk component-to-channel-invertible-outs (ntk)
  (forall (component i)
          (implies (and (componentp component (len (xmasnetwork-channels ntk)))
                        (< i (len (component-outs component))))
                   (equal (get-init-component (get-out-channel component i ntk) ntk)
                          component))))
\end{lstlisting}

For channels, we have a similar, but weaker, way of expressing the relationship between channels and components. Components can have multiple input/outputs. 
As it is unknown to the channel to which port of the component it is connected, we use two \func{defun-sk} to express the weaker property
that each component connected to a channel must have that channel connected to one of the ports. We specify this for both input and output ports of components.

\begin{lstlisting}[numbers=none]
(defun-sk component-has-channel-as-input (component channel ntk)
  (exists (i)
          (and (natp i)
               (< i (len (component-ins component)))
               (equal (get-in-channel component i ntk)
                      channel))))

(defun-sk channel-to-component-invertible-target (ntk)
  (forall (channel)
          (implies (channelp channel (len (xmasnetwork-components ntk)))
                   (and (componentp (get-target-component channel ntk) (len (xmasnetwork-channels ntk)))
                        (component-has-channel-as-input (get-target-component channel ntk)
                                                        channel
                                                        ntk)))))
\end{lstlisting}

Likewise, we specified a relationship between the initiator part of a channel and the output port of a component.
%
These \func{defun-sk}{}'s also ensure that an output port of a component is always connected to an input port of another component.

The proofs in the next section are predicated on a syntactically correct \xmas network. 
We introduce a predicate recognising such correct \xmas networks. 
The predicate includes the previous predicates and additional ones stating that the lists of components and channels are lists and contain 
no duplicates\footnote{It should be possible to deduce that there are no duplicates in the list of channels by proof of contradiction with the relation between channels and components. A duplicate in the list would contradict the fact that the relationship is invertible. This proof is out of scope of this paper and left as an exercise for the reader.}.

\begin{lstlisting}[numbers=none]
(defun xmasnetworkp (ntk)
  (and (xmasnetwork-p ntk)
       (componentsp (xmasnetwork-components ntk) 
                    (len (xmasnetwork-channels ntk)))
       (channelsp (xmasnetwork-channels ntk) 
                  (len (xmasnetwork-components ntk)))
       (component-to-channel-invertible-ins ntk)
       (component-to-channel-invertible-outs ntk)
       (channel-to-component-invertible-init ntk)
       (channel-to-component-invertible-target ntk)
       (no-duplicatesp (xmasnetwork-channels ntk))
       (no-duplicatesp (xmasnetwork-components ntk))
       (true-listp (xmasnetwork-channels ntk))
       (true-listp (xmasnetwork-components ntk))))
\end{lstlisting}

\section{Formal semantics of an \xmas network}
\label{sec:semantics}

\subsection{A simple example}
\label{sec:example}

\begin{figure}[h]
\centering
\includegraphics[scale=0.75]{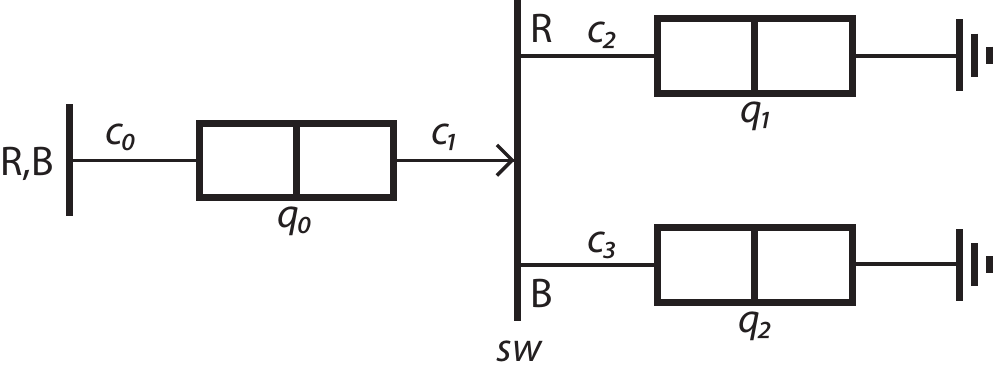}
\caption{Microarchitectural model of a small routing network which gets as input packets of type red and blue, and route them according to their type.}
\label{fig:xmas-red-blue}
\end{figure}

All \xmas primitives have well-defined semantics in the form of Boolean equations over channel signals. 
All equations are available in the paper introducing the \xmas language~\cite{chatterjee2010ieee}. 
As an example, the semantics of a switch primitive are given by 
the following equations, where $a$ and $b$ are the two output channels of a switch, and $i$ is the input channel. Function $s$ is used to determine 
the routing destination -- channel $a$ or channel $b$ -- depending on the content of the incoming packet:

\begin{eqnarray*}
	a.irdy & := & i.irdy \land s(i.data) \\
	b.irdy & := & i.irdy \land \lnot s(i.data) \\
	a.data & := & i.data \\
	b.data & := & i.data \\
	i.trdy & := & (a.\irdy \land a.\trdy) \lor (b.\irdy \land b.\trdy)
\end{eqnarray*}

Let us consider the network example in Figure~\ref{fig:xmas-red-blue}. 
This network is composed of a source injecting messages of type 'red' and 'blue' into 
queue $q_0$. A switch then routes red messages to $q_1$ and blue messages to $q_2$.
Red and blue messages are then sunk in two sinks. 

Initially, all queues are empty. Thus, all queues are ready to accept messages. The \trdy signal of their input channel is high. 
As they are empty, the \irdy signal of their output channel is low.
As explained earlier in Section~\ref{sec:xmas}, the execution semantics of \xmas proceed in two steps. 
First, all signals are updated. Second, queues are updated whenever transfers are possible. 
Assume the source emits a red message. The \irdy signal of its output channel is high. 
For channel $c_0$, both \irdy and \trdy are high. A transfer occurs and the red message moves 
to queue $q_0$. In the next combinatorial phase, the equations of the switch will propagate a high \irdy signal 
to channel $c_2$. The switch will also propagate a high $\trdy$ signal back to channel $c_1$. At the end of this 
combinatorial phase, channels $c_1$ and $c_2$ have their $\trdy$ and $\irdy$ signals set to high. Data will flow 
from $q_0$ to $q_1$. A similar computation will let the data flow to the sink. 

This simple example shows the dependency between routing and message content, but also the interdependency between $\irdy$ and $\trdy$ signals. 
These dependencies make the formalisation of the combinatorial part of the \xmas semantics not straightforward. 
The computation of \irdy signals depends on the computation of \trdy signals and \textit{vice versa}. 
Therefore there is no clear separation between routing and transfer decisions, as defined in the five functions of \genoc. 
Routing and transfer have to be computed simultaneously. 

In the remainder of this section, we define function \func{xmas-transfer-calculate} in such a way that it closely respects the \xmas semantics and 
can be used to instantiate functions routing and transfer of the \genoc definition.

\subsection{Datastructures}

Function \func{xmas-transfer-calculate} computes three elements. First, it computes the Boolean value of a signal. 
Then, it computes a set of routing targets and a set of transfer destinations. 
The former is used to match the routing phase of \genoc. The later is used to match the transfer 
phase of \genoc. 
Concretely the data structure used to calculate the result of the \irdy and \trdy wires is a triple $\tup{b,r,t}$, 
where $b$, $r$, and $t$ are 
defined as follows:
\begin{itemize}
\item $b$, signal on the wire. Boolean value of the wire, used during the computation of the transfer destinations. This value is actually on the \irdy and \trdy wire.

\item $r$, routing destinations. Routing destination $r$ is list of pairs (queue,packet). 
Each pair specifies a possible destination queue. The packet part specifies the actual content of the message. 
Content may change in the presence of function primitives on the path between two queues. 

\item $t$,  transfer destinations. Transfer destination $t$ has the same form as routing destination $r$. 
The difference is that routing destinations are \emph{potential} destinations, while transfer destinations 
are destinations which are actually reachable under the current network state. 

\end{itemize}

The \func{defstructure} definition of this structure is listed below. To create this structure we have created a wrapper, called \func{xmas-result} to specifically deal with a possible error state.
\begin{lstlisting}[numbers=none]
(defstructure xmas bool routing transfer)

(defun xmas-result (wire-signal routing transfer)
  (if (equal wire-signal 'error)
    'error
    (xmas wire-signal routing transfer)))
\end{lstlisting}

\subsection{Semantics of the \xmas primitives}

The semantics of each primitive are expressed over triples $\tup{b,r,t}$ instead of wires. 
We therefore need special conjunction and disjunction operations over these triples.

Conjunction operation is defined as follows, where $x,a,b$ are triples $\tup{b,r,t}$:

\[
x \mbox{~xmas-and~} y :=~<x.b \land y.b, x.r \cup y.r, \func{if}~x.b~\land~y.b~\func{then}~x.t~\cup~y.t~\func{else}~\emptyset~\func{fi}>
\]

Routing is over approximated by the union of the routing targets. Transfer is defined as the union of both 
transfer fields if both Boolean are true. Otherwise, it becomes the empty set. The Boolean field is updated with 
the usual logical conjunction. 
In \acl this is represented by the following function:
\begin{lstlisting}[numbers=none]
(defun xmas-and (x y)
  (let ((routing (append (xmas-routing x) (xmas-routing y)))
    (cond ((or (equal x 'error) (equal y 'error)) 'error)
          ((and (xmas-bool x) (xmas-bool y))
           (xmas-result t routing (append (xmas-transfer x) (xmas-transfer y))))
          (t (xmas-result nil routing nil)))))
\end{lstlisting}

Disjunction operation 
\emph{xmas-or} is defined as follows,  where $x,a,b$ are triples $\tup{b,r,t}$: 

\[
x \mbox{~xmas-or~} y :=~<x.b \lor y.b, x.r \cup y.r, (\func{if}~x.b~\func{then}~x.t~\func{else}~\emptyset~\func{fi}) \cup (\func{if}~y.b~\func{then}~y.t~\func{else}~\emptyset~\func{fi})>
\]

Here again, routing is over approximated with the union of the routing targets. The Boolean fields of $a$ and $b$ are used to select either 
the transfer destinations of $a$, or the transfer destinations of $b$, or the union of both transfer destinations. If the two Boolean fields are false, 
transfer destinations are empty. 
In \acl this is represented by the following function:
\begin{lstlisting}[numbers=none]
(defun xmas-or (x y)
  (let ((routing (append (xmas-routing x) (xmas-routing y)))
        (xt (xmas-transfer x))
        (yt (xmas-transfer y))
        (xb (xmas-bool x)) 
        (yb (xmas-bool y)))
    (cond ((or (equal x 'error) (equal y 'error)) 'error)
          ((and xb yb) (xmas-result t routing (append xt yt)))
          (xb (xmas-result t routing xt))
          (yb (xmas-result t routing yt))
          (t (xmas-result nil routing nil)))))
\end{lstlisting}

Encoding the semantics of each primitive is now a direct translation of the \irdy, \trdy, and \data equations. 
For instance, the following \trdy equation of the input channel of a switch: 

  \[
i.trdy := (a.\irdy \land a.\trdy) \lor (b.\irdy \land b.\trdy)
\]

is directly translated to the following ACL2 term:

\begin{lstlisting}[numbers=none]
(xmas-or
  (xmas-and
    (xmas-transfer-calculate 'irdy (get-out-channel c 0 ntk) ntk unvisited ntkstate)
    (xmas-transfer-calculate 'trdy (get-out-channel c 0 ntk) ntk unvisited ntkstate))
  (xmas-and
    (xmas-transfer-calculate 'irdy (get-out-channel c 1 ntk) ntk unvisited ntkstate)
    (xmas-transfer-calculate 'trdy (get-out-channel c 1 ntk) ntk unvisited ntkstate))))
\end{lstlisting}

where function \func{xmas-transfer-calculate} (see next section) actually computes the values of the triples associated to each signal. 
All of the five primitives -- queue, switch, source, sink, and function -- currently supported are translated in a similar way.

\subsection{Implementation of \func{xmas-transfer-calculate}}

Function \func{xmas-transfer-calculate}, see Figure~\ref{xtc}, computes the value of a specific signal on a given channel.
Argument {\tt flg} indicates which signal (\irdy, \trdy, or \data) is computed, on the channel specified by argument {\tt channel}.
Input argument {\tt unvisited} is a list of pairs {\tt (channel flg)}, which is initially all the pairs in the network. It is used to remember which signals have not already been computed and guarantee termination, which is explained later. 
Extra arguments to the function are the \xmas network and the current network state, in which the current contents of all the resources are stored. Both extra arguments are not changed during the computation. They are not returned in the result.

Note that the functions \func{xmas-can-send} and \func{xmas-can-receive} express the conditions under which a queue is ready to send 
a packet and when a queue is ready to receive a packet. In brief, a queue is ready to send a packet if it stores at least one message. 
A queue is ready to receive a packet if it is not full. \func{xmas-can-send} returns a triple $\tup{b,r,t}$ with empty routing and target destinations. \func{xmas-can-receive} returns a triple $\tup{b,r,t}$ in which the routing destinations contain the current queue, and if and only if the queue can receive a packet the transfer destinations contain the current queue, otherwise it is empty.

\begin{figure}
\begin{lstlisting}[basicstyle=\ttfamily\footnotesize]
(defun xmas-transfer-calculate (flg channel ntk unvisited ntkstate)
 (declare (xargs :measure (len unvisited)))
 (cond ((not (member-equal (cons channel flg) unvisited)) 'error) ; combinatorial cycle
  ((equal flg 'data)
   (let* ((cpt (get-init-component channel ntk))
          (type (component-type cpt))
          (next-unvisited  (remove1 (cons channel flg) unvisited)))
    (cond
     ((equal type 'queue) (xmas-next-data cpt ntkstate))
     ((equal type 'switch)
      (xmas-transfer-calculate 'data (get-in-channel cpt 0 ntk) ntk next-unvisited ntkstate))
     ...)))
  ((equal flg 'irdy)
   (let* ((cpt (get-init-component channel ntk))
          (type (component-type cpt))
           (index-out (if (equal (get-in-channel cpt 0 ntk) channel) 0 1))
           (next-unvisited  (remove1 (cons channel flg) unvisited)))
    (cond
     ((equal type 'queue)
      (xmas-result (xmas-can-send cpt ntkstate) nil nil))
     ((and (equal type 'switch) (equal index-out 0))
      (xmas-and (xmas-transfer-calculate 'irdy (get-in-channel cpt 0 ntk) ntk next-unvisited ntkstate)
                (xmas-result
                  (xmas-switch-function
                    cpt
                    (xmas-transfer-calculate
                      'data
                      (get-in-channel cpt 0 ntk)
                      ntk next-unvisited ntkstate))
                  nil
                  nil)))
      ...)))
  ((equal flg 'trdy)
   (let* ((cpt (get-target-component channel ntk))
          (type (component-type cpt))
          (next-unvisited  (remove1 (cons channel flg) unvisited)))
   (cond
    ((equal type 'queue)
     (xmas-component
       (xmas-can-receive cpt ntkstate)
       cpt
       (xmas-transfer-calculate 'data channel ntk next-unvisited ntkstate)))
    ((equal type 'switch) 
     (xmas-or
       (xmas-and 
         (xmas-transfer-calculate
           'irdy 
           (get-out-channel cpt 0 ntk) 
           ntk next-unvisited ntkstate)    
         (xmas-transfer-calculate 'trdy (get-out-channel cpt 0 ntk) ntk next-unvisited ntkstate))                                   
       (xmas-and 
         (xmas-transfer-calculate 'irdy (get-out-channel cpt 1 ntk) ntk next-unvisited ntkstate)
         (xmas-transfer-calculate
           'trdy
           (get-out-channel cpt 1 ntk)
           ntk next-unvisited ntkstate))))
    ...)))
  (t 'error)))
\end{lstlisting}
\caption{Code listing of the \func{xmas-transfer-calculate} function.}
\label{xtc}
\end{figure}

Returning to the example presented 
earlier in Section~\ref{sec:example}, 
we illustrate the calculation of the $c_2.irdy$ signal by this function. 
The different steps of the calculation are given below, with \func{xtc} as abbreviation of \func{xmas-transfer-calculate}, and without the \func{ntk}, \func{unvisited} and \func{ntkstate} arguments for reasons of clarity. As in the example, the packet in front of queue $q_0$ is of type \emph{red}.
This packet will be switched to top most channel $c_2$.
%

\begin{lstlisting}[numbers=none]
(xtc 'irdy c2) = (xmas-and (xtc 'irdy c1)
                           (xmas-switch sw (xtc 'data c1)))
               = (xmas-and (xmas-result (xmas-can-send q0) nil nil)
                           (xmas-switch sw 'red))
               = (xmas-and (xmas t nil nil)
                           (xmas t nil nil))
               = (xmas t nil nil)
\end{lstlisting}

A trace of the calculation of the signal on wire $c_1.trdy$ is stated below. Note that \func{xmas-not} can have three values as input, namely \func{true}, \func{false} and \func{error}. It preserves the \func{error} symbol, but otherwise functions like a normal \func{not}.

\begin{lstlisting}[numbers=none]
(xtc 'trdy c1) = (xmas-or (xmas-and (xtc 'irdy c2)
                                    (xtc 'trdy c2))
                          (xmas-and (xtc 'irdy c3)
                                    (xtc 'trdy c3)))
               = (xmas-or (xmas-and (xmas-and (xtc 'irdy c1)
                                              (xmas-result
                                                (xmas-switch-function sw (xtc 'data c1))
                                                nil nil))
                                    (xmas-can-receive q1))
                          (xmas-and (xmas-and (xtc 'irdy c1)
                                              (xmas-result
                                                (xmas-not
                                                 (xmas-switch-function sw (xtc 'data c1)))
                                                nil nil))
                                    (xmas-can-receive q2)))
               = (xmas-or (xmas-and (xmas-and (xmas-can-send q0)
                                              (xmas-result
                                                (xmas-switch-function sw 'red)
                                                nil nil))
                                    (xmas-can-receive q1))
                          (xmas-and (xmas-and (xmas-can-send q0)
                                              (xmas-result
                                                (xmas-not (xmas-switch-function sw 'red))
                                                nil nil))
                                    (xmas-can-receive q2)))
               = (xmas-or (xmas-and (xmas-and (xmas t nil nil)
                                              (xmas t nil nil))
                                    (xmas t (list q1) (list q1)))
                          (xmas-and (xmas-and (xmas t nil nil)
                                              (xmas nil nil nil))
                                    (xmas t (list q2) (list q2)))
               = (xmas t (list q1) (list q1))
\end{lstlisting}

\subsection{Combinatorial cycles and termination of \func{xmas-transfer-calculate}}
A combinatorial cycle is a dependency cycle that occurs if a value on a (\irdy or \trdy) wire depends on the value of itself. This is an undesired situation, as the output is not deterministic and depends on the previous state, electrical fluctuations, and timing of the updating signals. The semantics of this situation is also not specified in the original paper on \xmas. We avoid this situation by remembering the visited wires.

By inverting the set of visited wires we can prove termination of \func{xmas-transfer-calculate}. Each recursive call to the function to calculate new values removes from the set of unvisited nodes the current node. With each recursion step the measure decreases. If the current node is not in the set of unvisited nodes, the value is already calculated and depends on its own value. In this case we return an error value. This error value is propagated by all the \xmas-and and \xmas-or operations.

We can only reason about these networks formally, if the semantics are fully specified. As combinatorial cycles are unspecified, we therefore have to exclude them, in case of our function by assuming the error state is never returned. It can be easily checked that the error state is never returned by calculating all the values in a given input network once.

\subsection{Properties proven}

We proved the four important proof obligations stated in Section~\ref{sec:genoc} for our formalisation of the \xmas semantics. 

\paragraph{There is always at least one routing target}
This proof obligation state that routing must produce at least one valid routing target. 
This means that the list of routing target computed by function \func{xmas-transfer-calculate}
must be non-empty. 
\begin{lstlisting}[numbers=none]
(defthm xmas-calculate-at-least-one
  (let ((result (xmas-transfer-calculate 'trdy channel ntk unvisited ntkstate)))
    (implies (and (xmasnetworkp ntk)
                  (member-equal channel (xmasnetwork-channels ntk))
                  (not (equal result 'error)))
             (consp (xmas-routing result)))))
\end{lstlisting}
The proof of this theorem requires a hint, uses about 10 lemmas, and needs about 65,000 prover steps. 

\paragraph{All routing targets are resources}
An obvious consistency requirement is that all routing destinations are in fact resources, i.e. components that can contain packets, and not other components like switches or functions.
To test for this condition we have created a function \func{A-resources} checking that 
all members of the input list are resources, as defined by constant {\tt resourcetypes}.
\begin{lstlisting}[numbers=none]
(defthm xmas-transfer-calculate-target-are-resources
  (let ((result (xmas-transfer-calculate signal channel ntk unvisited ntkstate)))
    (implies (and (xmasnetworkp ntk)
                  (member-equal signal '(irdy trdy))
                  (member-equal channel (xmasnetwork-channels ntk)))
             (A-resources (strip-cars (xmas-routing result))) ntk)))
\end{lstlisting}
The proof of this theorem requires a hint, uses about 15 lemmas, and needs about 2,100,000 prover steps.

\paragraph{Transfer is a subset of routing}
As discussed earlier, a central part of our approach is the modelling for both the routing and transfer phases. 
An important proof obligation of \genoc is that the destinations chosen in the transfer phase 
are a subset of the routing targets. This is expressed in the theorem listed below.
\begin{lstlisting}[numbers=none]
(defthm transfer-is-subset-of-routing
  (let ((result (xmas-transfer-calculate signal channel ntk unvisited ntkstate)))
    (implies (member-equal signal '(irdy trdy))
             (subsetp-equal (xmas-transfer result)
                            (xmas-routing result)))))
\end{lstlisting}
The proof of this theorem requires no hint, uses about 7 lemmas, and needs about 700,000 prover steps. 

\paragraph{All transfer result have available space in their queues}
An important constraints from \genoc is that all the results from the transfer phase have indeed space in their buffers to receive a packet. 
The theorem defined below specifies that all resources returned by the transfer phase can receive a packet, as determined by function \func{xmas-can-receive}.
\begin{lstlisting}[numbers=none]
(defthm xmas-transfer-implies-available-resource
  (let ((result (xmas-transfer-calculate 'trdy channel ntk unvisited ntkstate)))
    (implies
      (and (xmasnetworkp ntk)
           (member-equal channel (xmasnetwork-channels ntk))
           (member-equal resource (strip-cars (xmas-transfer result))))
      (xmas-can-receive resource ntkstate))))
\end{lstlisting}
The proof of this theorem requires a hint, uses about 10 lemmas, and needs about 500,000 prover steps. 


\section{Conclusion and Future Work}
\label{sec:conclusion}

We presented a formalisation of five of the eight primitives of the \xmas language. The formalisation is mainly 
achieved by the function \func{xmas-transfer-calculate}. This function computes Boolean values of control wires, 
a list of potential routing targets, and a list of actual destinations based on the network state. The construction of this function allows 
for a possible integration within the \genoc environment. The later also provides an easy way to model the 
combinatorial and the sequential parts of the \xmas semantics. We discharged important proof obligations 
on the function \func{xmas-transfer-calculate}, in particular, non-emptiness of routing and correctness of transfer 
decisions. These proof obligations will be key to the integration into \genoc. 

There are three primitives that still need to be formalised. 
The formalisation of the fork primitive will require an extension of function \func{xmas-transfer-calculate}.
This modification will be needed to distinguish a choice between two routing targets 
from the necessity to route to two different targets simultaneously. 
Instead of simply gathering routing targets in a list, we probably need expressions made of conjunctions 
and disjunctions of routing targets. 
Merges and joins will impact the ordering and switching functions of \genoc. 
Their formalisation will require modification of the main \genoc function. 
In this direction, the next step is to extend \genoc with the possibility of modelling message dependent routing. 
This requires a re-definition of the notion of correctness. When routing is only based on the destination 
and the current position of a message, the notion of the correct destination is obvious. When routing 
depends on the content of a message, and functions may change this content on the way, it is no more 
obvious which destination is the correct one. There are many more challenges on the way towards 
a full formalisation of \xmas. The result will be a solid environment for the verification of detailed micro-architectural 
models of communication fabrics and a strong theory on communication networks architectures.


\bibliographystyle{eptcs}
\bibliography{../common}
\end{document}

%% file: common.tex
\usepackage{color}

\newcommand{\todo}[1]{\emph{\color{red} #1}}

\usepackage{xspace}

\usepackage{fancyvrb}
\newcommand{\dutch}{
	\setlength{\parskip}{1.3ex plus 0.2ex minus 0.2ex}
	\setlength{\parindent}{0pt}
	}

\definecolor{dkgreen}{rgb}{0,0.6,0}
\definecolor{grey}{rgb}{0.5,0.5,0.5}
\usepackage{listings}
\lstdefinelanguage{acl2}{
morekeywords={defun,defcong,defthm,t,nil,declare,defspec,local,defstructure},
keywordsprefix=:,
otherkeywords={defun-sk,rule-classes,in-theory,begin-book,include-book,in-package,mutual-recursion,instance-of-defspec},
sensitive=false,
alsodigit=-,
showstringspaces=false,
breaklines=true,
breakatwhitespace=true,
morecomment=[l];,
morecomment=[s]{\#|}{|\#},
morestring=[b]",
stringstyle=\color{dkgreen}}

\newcommand\genoc[0]{{\sc GeNoC}\xspace}
\newcommand\acl[0]{{\sc ACL2}\xspace}

\newcommand\xmas[0]{{\sc xM\!\!\;A\!\!\;S}\xspace}

\newcommand{\irdy}{\ensuremath{\mathit{irdy}}\xspace}
\newcommand{\trdy}{\ensuremath{\mathit{trdy}}\xspace}
\newcommand{\data}{\ensuremath{\mathit{data}}\xspace}

\newcommand\func[1]{{\sf #1}\xspace} 
\newcommand{\tup}[1]{\langle #1 \rangle} 